\documentclass[showpacs,showkeys,amssymb,aps,epsfig,graphicx,graphics,epsf]{revtex4}
\newcommand{\be}{\begin{equation}} \newcommand{\ee}{\end{equation}}
\newcommand{\bea}{\begin{eqnarray}}\newcommand{\eea}{\end{eqnarray}}

\usepackage{graphicx}

\begin{document}
\hfill{SINP/TNP/2008/16}
\vspace*{10mm}
\title{Bound States in Gapped Graphene with Impurities : Effective Low-Energy Description of Short-Range Interactions}
\author{Kumar S. Gupta\footnote{Email : kumars.gupta@saha.ac.in~ }}

\affiliation{ Theory Division, Saha Institute of Nuclear Physics, 1/AF Bidhannagar, Calcutta - 700064, India }

\author{Siddhartha Sen \footnote{Email : sen@maths.ucd.ie, Emeritus Fellow, TCD}}

\affiliation {School of Mathematical Sciences, UCD, Belfield, Dublin 4, Ireland}
\affiliation {Department of Theoretical Physics, Indian Association for the Cultivation of Science, Calcutta - 700032, India }

\begin{abstract}

We obtain a novel bound state spectrum of the low energy excitations near the Fermi points of gapped graphene in the presence of a charge impurity. The effects of possible short range interactions induced by the impurity are modelled by suitable boundary conditions. The spectrum in the subcritical region of the effective Coulomb coupling is labelled by a parameter which characterizes the boundary conditions and determines the inequivalent quantizations of the system. In the supercritical region we obtain a renormalization group flow for the effective Coulomb coupling.

\end{abstract}

\pacs{03.65.Ge, 81.05.Uw }
\keywords{Bound states, Graphene}
\maketitle

\section{Introduction}

Graphene monolayers consist of carbon atoms arranged in a honeycomb lattice. In the neighbourhood of Fermi points, the low energy excitations in graphene can be described by a two dimensional massless Dirac equation \cite{sem}. The effects of charge impurities in graphene have to be analyzed separately in two regions, subcritical and supercritical, depending on the strength of the Coulomb interaction. In the supercritical region, where the effective Coulomb strength exceeds a certain critical value, the massless Dirac equation admits bound states \cite{castro,levi1,levi2,us}. In the subcritical region this does not happen, which is a manifestation of the Klein paradox.

If the exact honeycomb lattice symmetry in graphene is partially broken, possibly due to the presence of an impurity, a Dirac mass for the excitations can be generated. The effect of a charge impurity in gapped graphene with massive Dirac excitations has been analyzed in \cite{ho,novi,kot1,kot2}, where it was assumed that the impurity provides an axially symmetric Coulomb interaction. It is expected that such an impurity may also induce other short-range or singular interactions, such as a delta function type potential. We have neither any detailed knowledge of such interaction terms, nor is it practical to include them in the Dirac Hamiltonian, which is valid only in the long wavelength limit. We can however still model the combined effect of these additional short range interactions on the long wavelength dynamics through the choice of suitable boundary conditions \cite{mol}. 

In this paper we shall analyze the effects of these boundary conditions on the spectrum of the massive Dirac equation in the presence of a charge impurity.  This approach, where self-adjointness is taken as the guiding principle for determining the allowed boundary conditions, has been shown by Jackiw to yield a reliable description singular potential such as a  delta function \cite{jackiw}. For instance, application of this approach to singular short range interactions has led to the possibility of novel bound states in fermionic \cite{ld1,ld2,ld3,ld4,ld5} and anyonic systems \cite{an1,an2}, molecular physics \cite{mol} and integrable models \cite{biru1,biru2}. This technique is particularly relevant for systems having scaling interactions  \cite{sae1, sae2, sae3, sae4}, a property present in the screening effect of the Coulomb potential in graphene \cite{screen, screen1, screen2, screen3, screen4}. Moreover, this approach has already been used to study certain  topological defects in graphene \cite{topo1,topo2}. It is thus of interest to explore the physical effects of generalized boundary conditions in gapped graphene with a Coulomb charge impurity. 

It should be noted that the analysis of the bound states in \cite{ho,novi} assumes that the wavefunctions vanish at the location of the impurity, whereas we only require square-integrabilty for the wave functions. The boundary conditions that are consistent with this requirement introduce an extra parameter, which has physical implications. In particular, the result for the spectrum obtained here is in general different from that obtained in \cite{ho,novi}. The parameter which appears in the boundary condition characterizes the inequivalent sectors of the quantum theory. The appropriate theoretical description for graphene in the presence of a charge impurity can be settled experimentally, possibly through the STM measurements of the local density of states (LDOS). 

It is generally believed that the long wavelength Dirac description is not applicable to graphene in the supercritical region, where the Dirac vacuum is expected to break down \cite{novi}. However certain features of the system may still be captured by the continuum description. For instance, numerical \cite{castro} as well as semi-classical analysis \cite{levi2} of the supercritical region in the massless case exhibits a large number of bound states in graphene. An analytical prediction of these bound states can be obtained from the continuum massless Dirac description \cite{us}. Here we shall apply the Dirac picture to the supercritical region with a cutoff comparable to the lattice spacing. We propose a renormalization group analysis by keeping the observables fixed as a function of the cutoff \cite{jackiw, rajeev}, to evaluate the corresponding $\beta$-function. An alternative regularization scheme for analyzing supercritical charge impurities in graphene with a mass gap has been discussed in \cite{kot1}. It is not necessary to introduce such a cutoff in the subcritical region as the excitation energies are low compared to the inverse lattice scales.

This paper is organized as follows. In Section 2, we set up the Dirac equation for the problem. In Section 3 we discuss the generalized boundary conditions that follow from self-adjointness. In Section 4 we find the spectrum with these boundary conditions. In Section 5 we discuss the spectrum in the supercritical region and the associated renormalization group flow. We conclude the paper in Section 6.

\section{Dirac Equation}

We start by considering the massive Dirac equation in a gapped graphene monolayer in the presence of a charge impurity and use the same conventions as in Novikov \cite{novi}. The Dirac operator can be written as
\be \label{op}
H = -i(\sigma_1 \partial_x + \sigma_2 \partial_y) + m \sigma_3 + V(r),
\ee
 where $r$ is the radial coordinate on the two dimensional $x-y$ plane, $\sigma_i,~i=1,2,3$ are the Pauli matrices, and $m$ denotes the Dirac mass. We have chosen units such that the Fermi velocity $v=1$ and the Planck's constant $\hbar = 1$. Using these conventions, the Coulomb potential $V(r)$ is given by 
\be
V(r) = - \frac{\alpha}{r},
\ee
where we choose the impurity strength $\alpha >0$, signifying an attractive potential \cite{screen}. In addition, we assume that the effect of the charge impurity is such that it induces short range and possibly singular potentials, such as a delta function, whose detailed nature is not relevant. In our approach, we assume that the combined effect of these short range and possibly singular potentials can be modelled by imposing suitable boundary conditions on the wave function.

The Dirac operator (\ref{op}) satisfies the eigenvalue equation
\be \label{Deq}
H \Psi = E \Psi
\ee
where $E$ is the eigenvalue and
\be \label{separ} 
\Psi (r, \phi) = \left( \begin{array}{c}
{\psi_1 (r ) ~ \Phi_k(\phi) } \\
{i \psi_2 (r )~ \Phi_{k+1}(\phi) } \\
\end{array} \right), ~~~
\Phi_k(\phi) = \frac{1}{\sqrt{2 \pi}} e^{ik\phi}, ~~~ k \in Z,
\ee
Here $\psi_1 (r)$ and $\psi_2 (r)$ denote the radial part of the wavefunction and $\phi$ denotes the angle in the $x-y$ plane. 

In this paper we focus on the bound states of the Dirac equation (\ref{Deq}), which satisfy  $|E| < m$. Consider the ansatz
\bea \label{ans}
\psi_1 (\rho) &=& \sqrt{m + E} e^{-\frac{\rho}{2}} \rho^{\nu - \frac{1}{2}} \chi_1 (\rho), \\
\psi_2 (\rho) &=& \sqrt{m - E} e^{-\frac{\rho}{2}} \rho^{\nu - \frac{1}{2}} \chi_2 (\rho),
\eea
where $\rho = 2 \gamma r$, $\gamma = \sqrt{m^2 - E^2}$, $\nu = \sqrt{j^2 - \alpha^2}$ and $j = k +\frac{1}{2}$. In this Section we shall deal with the subcritical region of the Coulomb potential, which is given by $\alpha < j$ for any $j$. Since the lowest value of $j = \frac{1}{2}$, in the subcritical region the effective Coulomb strength must satisfy $ \alpha < \frac{1}{2}$. Furthermore, in terms of the variables $P, Q$ defined by
\be
\chi_1 = P + Q, ~~~ \chi_2 = P - Q
\ee
we get the equations
\be \label{rad}
H_\rho \left( \begin{array}{c}
P  \\
Q \\
\end{array} \right) =
\left( \begin{array}{cc}
\rho \frac{d}{d \rho} + \nu - \frac{\alpha E}{\gamma} & - j + \frac{m \alpha}{\gamma}  \\
- j - \frac{m \alpha}{\gamma} & \rho \frac{d}{d \rho} + \nu - \rho + \frac{\alpha E}{\gamma} \\
\end{array} \right)
\left( \begin{array}{c}
P  \\
Q  \\
\end{array} \right) = 0,
\ee
which $H_\rho$ defined above denotes the radial Dirac operator. These set of equations can be decoupled to give
\bea \label{hyper}
\rho \frac{d^2 P}{d \rho^2} + (1 + 2 \nu -\rho)\frac{d P}{d \rho} - \left ( \nu - \frac{\alpha E}{\gamma} \right )P &=& 0 \\
\rho \frac{d^2 Q}{d \rho^2} + (1 + 2 \nu -\rho)\frac{d Q}{d \rho} - \left (1 + \nu - \frac{\alpha E}{\gamma} \right )Q &=& 0 
\eea
Thus we see that the functions $P$ and $Q$ satisfy the confluent hypergeometric equation \cite{as}. This equation has two linearly independent solutions, one of which is regular at the origin (denoted by M) while the other is regular at infinity (denoted by U).
In \cite{novi}, the boundary conditions were so chosen that the solutions were regular at the origin, which led to the wavefunctions
\bea
\psi_1 (\rho) &=&  \sqrt{m + E} e^{-\frac{\rho}{2}} \rho^{\nu - \frac{1}{2}} 
[ (j + \frac{m \alpha}{\nu}) M(\nu - \frac{\alpha E}{\gamma}, 1+ 2\nu, \rho) + (\nu - \frac{\alpha E}{\gamma})
M(1 + \nu - \frac{\alpha E}{\gamma}, 1+ 2 \nu , \rho)] \\
\psi_2 (\rho) &=&  \sqrt{m - E} e^{-\frac{\rho}{2}} \rho^{\nu - \frac{1}{2}} 
[ (j + \frac{m \alpha}{\nu}) M(\nu - \frac{\alpha E}{\gamma}, 1+ 2\nu, \rho) - (\nu - \frac{\alpha E}{\gamma})
M(1 + \nu - \frac{\alpha E}{\gamma}, 1+ 2 \nu , \rho)].
\eea
The corresponding bound state spectrum was obtained in \cite{ho,novi} as
\be \label{spec}
E_{p,j} = \frac{m ~{\mathrm{sgn}} (\alpha)}{\sqrt{1 + \frac{\alpha^2}{(p + \nu)^2}}},
\ee
with $p = 0,1,2,...,$ for $j > 0$ and $p = 1, 2, 3, ..., $ for $j < 0$. In the next Section we shall see that more general boundary conditions are possible which are consistent with all the requirements of quantum mechanics, leading to a different spectrum for the same Dirac operator.

\section{Generalized Boundary Conditions}

In the usual description of quantum mechanics, it is assumed that the Hamiltonian is self-adjoint \cite{reed}, so that the time evolution is unitary and the probabilities are conserved. In addition, for the bound states, the solutions should be square-integrable. In our search for the generalized boundary conditions, we shall be guided by these principles as formulated by von Neumann \cite{reed}. 

The Dirac operator $H$ in (\ref{Deq}) consists of a radial and an angular part. The domain $Y(\phi)$ on which the angular part of $H$ acts is spanned by the periodic functions $\Phi_k(\phi)~, k \in Z$ in (4). In what follows, we shall leave the angular wavefunctions and the corresponding boundary conditions unchanged. 

The radial part of the Dirac operator $H$ is given by $H_\rho$ in (\ref{rad}). It is symmetric (or Hermitian) in the domain $D_0(H_\rho) = C_0^\infty(R^+)$ consisting of infinitely differentiable functions of compact support in the half line $R^+$. The corresponding adjoint operator is denoted by $H^{\dagger}_\rho$, which, as a differential operator, has the same expression as $H_\rho$ in (\ref{rad}), although its domain could be different. 

The domain $D_0(H)$ of the full Dirac operator $H$ is therefore given by
$D_0(H) = C_0^\infty(R^+) \otimes Y(\phi)$. Its adjoint operator $H^{\dagger}$ has the same differential expression as $H$, although its domain could be different as well.
Following von Neumann's approach \cite{reed}, in order to determine whether the full Dirac operator $H$ is self-adjoint in its domain $D_0(H)$, we consider the equation
\be \label{def}
H^{\dagger} \Psi_{\pm} = \pm i \Psi_{\pm}.
\ee
 Let $n_+(n_-)$ be the total number of square-integrable, linearly independent solutions  of (\ref{def}) with the upper (lower) sign in the right
hand side. The quantities $n_\pm$ are called the deficiency indices of
$H$.  In order to determine $n_+(n_-)$, we consider the radial equation (\ref{rad}) with E replaced by $+i(-i)$. This will give the deficiency indices $n_{\pm}$ for $H_\rho$. 
In terms of $n_{\pm}$, $H_\rho$ can be classified as follows \cite{reed} :\\ 1) $H_\rho$ is (essentially) self-adjoint in $D_0(H_\rho)$ iff $( n_+ , n_- ) = (0,0)$.\\ 2) $H_\rho$ is not
self-adjoint in $D_0(H_\rho)$ but admits self-adjoint extensions iff $n_+ = n_- = n (say)
\neq 0$.\\ 3)  $H_\rho$ has no self-adjoint extensions if $n_+ \neq n_-$.\\

In order to find the deficiency indices $n_{\pm}$ for $H_\rho$, we need to solve for $P_{\pm}$ and $Q_{\pm}$ from the equations
\bea \label{def1}
\rho \frac{d^2 P_{\pm}}{d \rho^2} + (1 + 2 \nu -\rho)\frac{d P_{\pm}}{d \rho} - \left ( \nu - \frac{\alpha i}{\gamma_{\pm}} \right )P_{\pm} &=& 0 \\
\rho \frac{d^2 Q_{\pm}}{d \rho^2} + (1 + 2 \nu -\rho)\frac{d Q_{\pm}}{d \rho} - \left (1 + \nu - \frac{\alpha i}{\gamma_{\pm}} \right )Q_{\pm} &=& 0,
\eea
which are obtained from (9) and (10) with E replaced everywhere with $\pm i$ and where $\gamma_{\pm} = \sqrt{M^2 + 1}$. The solutions we seek are such that when we reconstruct 
\be
\chi_{1\pm} = P_\pm + Q_\pm, ~~~ \chi_{2\pm} = P_\pm - Q_\pm
\ee
and subsequently obtain
\bea \label{anspm}
\psi_{1\pm}  &=& \sqrt{m \pm i} e^{-\frac{\rho}{2}} \rho^{\nu - \frac{1}{2}} \chi_{1\pm} (\rho), \\
\psi_{2\pm}  &=& \sqrt{m \mp i} e^{-\frac{\rho}{2}} \rho^{\nu - \frac{1}{2}} \chi_{2\pm} (\rho),
\eea
the functions $\psi_{1\pm}$ and $ \psi_{2\pm}$ would be square integrable on $R^+$ with a measure $\rho d \rho$.

We now proceed to find $n_+$. In this case, a possible set of solutions of (15) and (16) are given by
\bea
P_+ &=& U \left ( \nu - \frac{i \alpha}{\gamma_+}, 1 + 2 \nu, \rho \right ), \\
Q_+ &=& U \left ( 1 + \nu - \frac{i \alpha}{\gamma_+}, 1 + 2 \nu, \rho \right ),
\eea
where $U$ denotes a confluent hypergeometric function \cite{as}. As $\rho \longrightarrow \infty$,
\bea
P_+ &\longrightarrow& \rho^{-\nu + \frac{i \alpha}{\gamma_+}}\\
Q_+ &\longrightarrow& \rho^{-1 -\nu + \frac{i \alpha}{\gamma_+}}.
\eea
Using (17), (18), (19), (22) and (23) we find that as $\rho \longrightarrow \infty$,
$\psi_{1+}, \psi_{2+} \longrightarrow 0$. Hence the functions $\psi_{1+}, \psi_{2+}$ are square integrable at infinity.

Let us now consider the behaviour of these functions as $\rho \longrightarrow 0$. For this we shall use the formula \cite{as}
\be \label{asym}
U(a,b,z) = \frac{\pi}{\sin \pi b} \left [ \frac{M(a,b,z)}{\Gamma(1+a-b)\Gamma(b)} 
- z^{1-b} \frac{M(1+a-b,2-b,z)}{\Gamma(a)\Gamma(2-b)} \right ],
\ee
where as $\rho \longrightarrow 0$, $M(a,b,z) \longrightarrow 1$. Using (20) and (\ref{asym}) we see that as $\rho \longrightarrow 0$,
\bea
P_+ &\longrightarrow& a (A_+ - B_+ \rho^{-2 \nu}), \\
Q_+ &\longrightarrow& a (C_+ - D_+ \rho^{-2 \nu}),
\eea
where $a = \frac{\pi}{\sin \pi (1+ 2 \nu)} $ and
\bea
A_+ &=& \frac{1}{\Gamma(-\nu - \frac{i \alpha}{\gamma_+}) \Gamma(1 + 2 \nu)} ~~~~~~~
B_+ = \frac{1}{\Gamma(\nu - \frac{i \alpha}{\gamma_+}) \Gamma(1 - 2 \nu)} \\
C_+ &=& \frac{1}{\Gamma(1 -\nu - \frac{i \alpha}{\gamma_+}) \Gamma(1 + 2 \nu)} ~~~~
D_+ = \frac{1}{\Gamma(1 + \nu - \frac{i \alpha}{\gamma_+}) \Gamma(1 - 2 \nu)} 
\eea
are constants depending on the system parameters. From the above relations we find that as 
$\rho \longrightarrow 0$, 
\bea 
\int |\psi_{1+}|^2 \rho d \rho &\longrightarrow& \int (c_1 \rho^{2 \nu} + c_2 + c_3 \rho^{-2 \nu} ) d \rho \\
\int |\psi_{2+}|^2 \rho d \rho &\longrightarrow& \int (d_1 \rho^{2 \nu} + d_2 + d_3 \rho^{-2 \nu} ) d \rho, 
\eea
where $c_i, d_i, i=1,2,3$ are constants whose explicit forms are not relevant. Recall that $\nu = \sqrt{j^2 - \alpha^2}$ and that in the subcritical region, $\alpha < j$. Hence $\nu$ is a real positive quantity in the subcritical region. Then, from (29) and (30) we find that $\psi_{1+}, 
\psi_{2+}$ are square integrable at the origin provided $\nu < \frac{1}{2}$. Thus we arrive at the conclusion that the functions $\psi_{1+}, \psi_{2+}$ are square integrable everywhere provided
$0 < \nu < \frac{1}{2}$. Alternately we can say that the deficiency index $n_+ = 1$ when $0 < \nu < \frac{1}{2}$. A similar analysis shows that for this same range of $\nu$, $n_- = 1$ as well. 
We have thus shown that when $0 < \sqrt{j^2 - \alpha^2} < \frac{1}{2}$, the massive Dirac operator for graphene in the subcritical region of the effective Coulomb coupling is not self-adjoint in $D_0(H_\rho)$, but admits a one parameter family of self-adjoint extensions.

\section{Inequivalent Spectra}

We would now like to find the spectrum of the system in the range of $j$ and the effective subcritical Coulomb strength $\alpha$ such that $0 < \nu = \sqrt{j^2 - \alpha^2} < \frac{1}{2}$, where the Dirac operator admits a one-parameter family of self-adjoint extensions. The deficiency subspaces for the radial Dirac operator $H_\rho$ are spanned by the elements
\be
\eta_{\pm} = 
\left( \begin{array}{c}
\psi_{1 \pm}  \\
\psi_{2 \pm} \\
\end{array} \right) =
\left( \begin{array}{c}
\sqrt{m \pm i}~ e^{-\frac{\rho}{2}} \rho^{\nu - \frac{1}{2}} (P_\pm + Q_\pm)  \\
\sqrt{m \mp i}~ e^{-\frac{\rho}{2}} \rho^{\nu - \frac{1}{2}} (P_\pm - Q_\pm) \\
\end{array} \right).
\ee
The domain in which the Dirac operator is self-adjoint is then given by 
$D_z(H_\rho) = D_0(H_\rho) \oplus \{ c(e^{i z \over 2}\eta_+  + e^{-{i z \over 2}} \eta_- ) \}$ where $c$ is an arbitrary complex number and $z \in R$ mod $2 \pi$ \cite{reed}. Thus we have a one parameter family of self-adjoint extensions, labeled by a real parameter $z$. For each choice of the parameter $z$, we have a domain of self-adjointness of the radial Dirac operator defined by $D_z(H_\rho)$. An arbitrary element $ \eta_z \in D_z(H_\rho)$ can be written as
\be \eta_z
 \left( \begin{array}{c}
\eta_{1 z}  \\
\eta_{2 z} \\
\end{array} \right) =
c\left( \begin{array}{c}
(e^{i z \over 2} \psi_{1 +}  + e^{-{i z \over 2}}\psi_{1 -}) \\
(e^{i z \over 2} \psi_{2 +}  + e^{-{i z \over 2}}\psi_{2 -}) \\
\end{array} \right).
\ee
 We note that as $\rho \longrightarrow 0$,
\be
\left( \begin{array}{c}
\eta_{1 z}  \\
\eta_{2 z} \\
\end{array} \right) 
\longrightarrow 
c\left( \begin{array}{c}
\sqrt{m + i}e^{i z \over 2}\rho^{\nu - \frac{1}{2}}(P_+ + Q_+) +
\sqrt{m - i}e^{-{i z \over 2}}\rho^{\nu - \frac{1}{2}}(P_- + Q_-) \\
\sqrt{m - i}e^{i z \over 2}\rho^{\nu - \frac{1}{2}}(P_+ - Q_+) +
\sqrt{m + i}e^{-i {z \over 2}}\rho^{\nu - \frac{1}{2}}(P_- - Q_-)
\end{array} \right),
\ee
where $P_-$ and $Q_-$ denote the complex conjugates of $P_+$ and $Q_+$ in (25) and (26) respectively.

We now proceed to find the spectrum of the system when the boundary conditions are governed by the domain $D_z(H_\rho)$. A solution of the physical eigenvalue problem can be written as
\be
\psi = N
\left( \begin{array}{c}
\sqrt{m + E}~ e^{-\frac{\rho}{2}} \rho^{\nu - \frac{1}{2}} (P + Q)  \\
\sqrt{m - E}~ e^{-\frac{\rho}{2}} \rho^{\nu - \frac{1}{2}} (P - Q) \\
\end{array} \right)
\ee
where the functions $P$ and $Q$ satisfy (9) and (10) respectively and $N$ denotes the normalization. Solutions of (9) and (10) that are square integrable at infinity are given by
\bea
P &=& U \left ( \nu - \frac{\alpha E}{\gamma}, 1+ 2 \nu, \rho \right ) \\
Q &=& U \left ( 1 + \nu - \frac{\alpha E}{\gamma}, 1+ 2 \nu, \rho \right ).
\eea
Using (24), (33) and (34) we see that in the limit as $\rho \longrightarrow 0$,
\bea
P &\longrightarrow& a(A - B \rho^{-2 \nu}) \\
Q &\longrightarrow& a(C - D \rho^{-2 \nu}) 
\eea
where $a = \frac{\pi}{\sin \pi (1+ 2 \nu)} $ and
\bea
A &=& \frac{1}{\Gamma(-\nu - \frac{E \alpha}{\gamma}) \Gamma(1 + 2 \nu)} ~~~~~~~
B = \frac{1}{\Gamma(\nu - \frac{E \alpha}{\gamma}) \Gamma(1 - 2 \nu)} \\
C &=& \frac{1}{\Gamma(1 -\nu - \frac{E \alpha}{\gamma}) \Gamma(1 + 2 \nu)} ~~~~
D = \frac{1}{\Gamma(1 + \nu - \frac{E \alpha}{\gamma}) \Gamma(1 - 2 \nu)} .
\eea
Hence, as $\rho \longrightarrow 0$,
\be
\psi  \longrightarrow a N
\left( \begin{array}{c}
\sqrt{m + E}~ [(A + C) \rho^{\nu - \frac{1}{2}} - (B+D) \rho^{-\nu - \frac{1}{2}}]  \\
\sqrt{m - E}~ [(A - C) \rho^{\nu - \frac{1}{2}} - (B-D) \rho^{-\nu - \frac{1}{2}}] \\
\end{array} \right)
\ee

The physical solution $\psi$ in (41) must belong to the domain of self-adjointness given by $D_z(H_\rho)$. In fact, behaviour of the elements of the domain $D_z(H_\rho)$ determine the boundary conditions for the system. If $ \psi \in D_z(H_\rho)$, then as  $\rho \longrightarrow 0$, the coefficients of $r^{\nu -\frac{1}{2}}$ and $r^{-\nu - \frac{1}{2}}$ in (33) and (41) must match. Comparing such terms and defining $\sqrt{m + i}(A_+ + C_+) = \xi_1 e^{i \theta_1}$ and $\sqrt{m + i}(B_+ + D_+) = \xi_2 e^{i \theta_2}$, we obtain
\be
\left ( \frac{\gamma^2}{1 + M^2} \right )^\nu \frac{A+C}{B+D} = 
\frac{\xi_1 \cos(\theta_1 + \frac{z}{2})}{\xi_2 \cos(\theta_2 + \frac{z}{2})}.
\ee
Using (39), (40) and (42) we get
\be
f(E) \equiv
\left ( \frac{\gamma^2}{1 + M^2} \right )^\nu
\frac{\Gamma(1-2\nu) \Gamma(1 + \nu - \frac{E \alpha}{\gamma}) (1 - \nu - \frac{E \alpha}{\gamma})}{\Gamma(1+2\nu) \Gamma(1 - \nu - \frac{E \alpha}{\gamma}) (1 + \nu - \frac{E \alpha}{\gamma})} = \frac{\xi_1 \cos(\theta_1 + \frac{z}{2})}{\xi_2 \cos(\theta_2 + \frac{z}{2})}.
\ee

\begin{figure}
\begin{center}
\includegraphics[width=7cm]{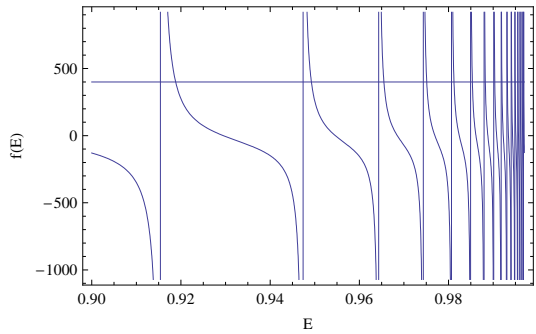}
\end{center}
\caption { \label{fig1} (Color online) A typical plot of $f(E)$ in (43) with $m=1$, $j = \frac{3}{2}$ and $\alpha = 1.46$. The horizontal line corresponds to the right hand side of (43). It can be shifted up or down by changing the self-adjoint extension parameter $z$.}
\end{figure}
Eqn. (43) determines the spectrum in terms of the system parameters and the self-adjoint extension parameter $z$. Each choice of $z$ corresponds to a boundary condition described by the domain $D_z(H_\rho)$ and leads to an inequivalent quantum theory. It may be noted that the theory itself cannot predict which choice of the self-adjoint extension parameter will be realized in a given system and this parameter must be determined empirically. Equation (43) in general cannot be solved analytically. However, for the special choice of $z = z_1$ such that $\theta_2 + \frac{z_1}{2} = \frac{\pi}{2}$, we have 
\be
\nu - \frac{E \alpha}{\gamma} = -n,~~~ n=1,2,3,....
\ee
This leads to the spectrum (\ref{spec}) obtained by Novikov \cite{ho,novi} for $0 < \nu < \frac{1}{2}$. For another special choice of $z = z_2$ such that $\theta_1 + \frac{z_2}{2} = \frac{\pi}{2}$, we get
\be
-\nu - \frac{E \alpha}{\gamma} = -n,~~~ n=1,2,3,....
\ee
For a general choice of $z$, the spectrum can be obtained numerically, and example of which is shown in Fig. 1. It may be noted that for a general choice of $z$, the spectrum we obtain from (43) is very different from that in (13), which was obtained previously \cite{ho,novi}. The corresponding bound state wavefunctions (41) are square-integrable, but not necessarily regular at the origin. This feature appears in graphene with topological defects as well \cite{topo1,topo2}.

\section{Supercritical region}

The supercritical region is defined by the effective Coulomb strength  $\alpha^2 > j^2$ for any $j$. This implies that in the supercritical region $\alpha > \frac{1}{2}$ and $\nu = \sqrt{j^2 - \alpha^2} = \pm i \mu$ where $\mu \in R$.  We now proceed to investigate the supercritical coupling region for the massive Dirac equation. A study  of the massive Dirac equation with a regularized Coulomb potential has been discussed in \cite{kot1}. We shall focus on the excitations satisfying $E^2 < m^2$ and introduce a cutoff in the radial direction set by the lattice spacing in graphene. The cutoff  restricts our analysis to the region where the Dirac equation holds. The corresponding  eigenvalue problem is solved with a hard-core boundary condition given by
\be
\psi (\rho=\rho_0) = 0,~~ \rho_0 = 2 r_0 \gamma,
\ee
where $\psi$ is the two component wavefunction in the supercritical region and $r_0$ provides a cutoff in the radial direction. In this case, the upper component $\psi_1(\rho)$ in (5) has two linearly independent solutions given by
\bea
\xi(\rho) &=& \sqrt{m + E} e^{-\frac{\rho}{2}}\rho^{i\mu - \frac{1}{2}}
M \left ( i\mu - \frac{\alpha E}{\gamma}, 1+ 2 i\mu, \rho \right ) \\
\zeta(\rho) &=& \sqrt{m + E} e^{-\frac{\rho}{2}}\rho^{-i\mu - \frac{1}{2}}
M \left ( -i\mu - \frac{\alpha E}{\gamma}, 1- 2 i\mu, \rho \right ).
\eea
The general solution which satisfies the boundary condition (46) can be written as
\be
\psi_1(\rho) =  [\xi(\rho) \zeta(\rho_0) - \zeta(\rho_) \xi(\rho_0)].
\ee
As $\rho \longrightarrow \infty$, we get that
\be
\psi_1(\rho) \longrightarrow \sqrt{m + E}e^{\frac{\rho}{2}}\rho^{-\frac{1}{2}}
\left [
\frac{\Gamma(1+ 2 i\mu)}{\Gamma(i\mu - \frac{\alpha E}{\gamma})} \zeta(\rho_0) 
- \frac{\Gamma(1- 2 i\mu)}{\Gamma(-i\mu - \frac{\alpha E}{\gamma})} \xi(\rho_0)
\right ].
\ee
In order for the wave function to be square integrable, the quantity in the parenthesis on the rhs of (50) must vanish. This gives the condition
\be
\frac{\Gamma(1+ 2 i\mu)\Gamma(-i\mu - \frac{\alpha E}{\gamma})}
{\Gamma(1- 2 i\mu)\Gamma(i\mu - \frac{\alpha E}{\gamma})} 
= \frac{\xi(\rho_0)}{\zeta(\rho_0)}.
\ee
Eqn. (51) follows as an exact consequence of our analysis. In order to gain some physical insight, we shall now use several approximations. The results derived below are therefore valid only in a qualitative fashion. First we assume that as the cutoff $r_0$ approaches the lattice spacing, the hypergeometric function $M$ in (47) and (48) can be replaced approximately by 1. Strictly speaking this is ture when the cutoff tends to zero, but it is a reasonable approximation in the long wavelength limit. Second, we assume that $E^2 < m^2$. In other words, we shall trust our results only for energy scales below the Dirac mass. Using these assumptions in (51), we get 
\be
\frac{\Gamma(-i\mu - \frac{\alpha E}{\gamma})}
{\Gamma(i\mu - \frac{\alpha E}{\gamma})} 
= e^{2 i (\mu \ln \rho_0 + \delta)} 
\ee
where $\delta$ is the argument of $\Gamma(1- 2 i\mu)$. In order to proceed, consider the energy scale such that $\frac{E}{m} << 1 $. In this case, the l.h.s. of (52) is approximately independent of $E$ and depends only on the system parameter $\mu$. Denoting the argument of $\Gamma(-i\mu)$ by $\theta$, we get
\be
\gamma_p  = \frac{1}{2r_0} e^\frac{\theta - \delta - 2 p \pi}{\mu },
\ee
where $p \in Z$ and $\gamma_p = \sqrt{m^2 - E_p^2}$. We can satisfy the requirement of $\frac{E}{m} << 1 $ by restricting $p$ suitably. We now assume that $\mu$, through its dependence on the effective Coulomb coupling $\alpha$, is a function of the cutoff $r_0$. We keep $E_p$ or equivalently $\gamma_p$ invariant as the cutoff is varied, which gives the $\beta$ function as
\be
\beta (\mu) = - r_0 \frac{d \mu}{d r_0} \sim - \mu^2.
\ee
We see that the coupling $\mu$ admits an ultraviolet stable fixed point at $\mu =0$ or equivalently at $\alpha = j$ for the angular momentum channel $j$, to which the system is expected to flow \cite{jackiw,rajeev}. In particular, $\alpha$ tends to its critical value $\frac{1}{2}$ for the angular momentum channel $j = \frac{1}{2}$.

\section{Conclusion}

In this paper we have used the freedom to choose generalized boundary conditions to model the effects of short range interactions introduced by impurities in gapped graphene. We used this approach to investigate the super and subcritical regions for the effective Coulomb charge. For the subcritical region we found that the generalized boundary conditions introduce a self-adjoint extension parameter $z$ which labels the different inequivalent quantizations for $0 < \sqrt{j^2 - \alpha^2} < \frac{1}{2}$. For a specific choice of $z$, the result of \cite{ho,novi} can be recovered. In general the spectrum obtained is different. Thus an experimental approach for determining the appropriate choice of the boundary conditions labelled by $z$ is in principle possible.

For the supercritical region, the analysis suggests a renormalization group flow $\alpha \rightarrow j$ for the $j^{\rm{th}}$ angular momentum channel, where $j$ is half integer. In particular, for $j=\frac{1}{2}$, the effective Coulomb coupling tends to its critical value  $\alpha=\frac{1}{2}$. This conclusion is valid in a very restricted region where $\frac{E}{m} << |\mu|$. 

In this paper we have considered only bound states. A similar analysis for the scattering sector would be relevant. In addition, the analysis of self-adjointness in bilayer graphene with impurities \cite{bi} would also be interesting.

\noindent
{\bf Acknowledgments}

We would like to thank S. Chakrabarti for the help with Mathematica, which has been used to generate Fig.1. We thank A. Samsarov for critically reading the manuscript and for kind comments.

\end{document}